\documentclass[letterpaper,twocolumn,prl,aps,superscriptaddress,amsmath,amssymb,floatfix]{revtex4-2}
\usepackage{mathptmx}
\usepackage{color}
\usepackage{amsmath}
\usepackage{amssymb}
\usepackage{graphicx}
\usepackage{esint}
\usepackage{siunitx}
\usepackage[unicode=true,
 bookmarks=true,bookmarksnumbered=false,bookmarksopen=false,
 breaklinks=false,pdfborder={0 0 1},backref=false,colorlinks=true]
 {hyperref}
\hypersetup{
 linkcolor=magenta,urlcolor=blue,citecolor=blue,pdfstartview={FitH},hyperfootnotes=false}

\makeatletter

\usepackage{textcomp}
\usepackage{epstopdf}

\usepackage{amsfonts}

\pdfpageheight\paperheight
\pdfpagewidth\paperwidth

\usepackage{xcolor}\usepackage{soul}
\newcommand{\ket}[1]{\ensuremath{\left|#1\right\rangle}}

\definecolor{blue}{rgb}{0,0,1}
\definecolor{red}{rgb}{1,0,0}
\definecolor{green}{rgb}{0,1,0}

\usepackage{soul}

\makeatother

\begin{document}

\title{Rydberg quantum antennas for chip-interfaced single-photon source array}

\author{Yan-Lei Zhang}
\thanks{These authors contribute equally to this work.}
\affiliation{Laboratory of Quantum Information, University of Science
and Technology of China, Hefei 230026, China}
\affiliation{Anhui Province Key Laboratory of Quantum Network, University of Science and Technology of China, Hefei 230026, China}
\affiliation{CAS Center For Excellence in Quantum Information and Quantum Physics,
University of Science and Technology of China, Hefei, Anhui 230026,
China}
\affiliation{Hefei National Laboratory, University of Science and Technology of China, Hefei 230088, China}

\author{Dong-Qi Ma}
\thanks{These authors contribute equally to this work.}
\affiliation{Laboratory of Quantum Information, University of Science
and Technology of China, Hefei 230026, China}
\affiliation{Anhui Province Key Laboratory of Quantum Network, University of Science and Technology of China, Hefei 230026, China}

\author{Guang-Jie Chen}
\affiliation{Fujian Provincial Key Laboratory of Quantum Manipulation and New Energy Materials, College of Physics and Energy, Fujian Normal University, Fuzhou 350117, China}

\author{Qing-Xuan Jie}
\affiliation{Laboratory of Quantum Information, University of Science
and Technology of China, Hefei 230026, China}
\affiliation{Anhui Province Key Laboratory of Quantum Network, University of Science and Technology of China, Hefei 230026, China}

\author{Liang Chen}
\affiliation{Laboratory of Quantum Information, University of Science
and Technology of China, Hefei 230026, China}
\affiliation{Anhui Province Key Laboratory of Quantum Network, University of Science and Technology of China, Hefei 230026, China}

\author{Ya-Dong Hu}
\affiliation{Laboratory of Quantum Information, University of Science
and Technology of China, Hefei 230026, China}
\affiliation{Anhui Province Key Laboratory of Quantum Network, University of Science and Technology of China, Hefei 230026, China}

\author{Zhu-Bo Wang}
\affiliation{Laboratory of Quantum Information, University of Science
and Technology of China, Hefei 230026, China}
\affiliation{Anhui Province Key Laboratory of Quantum Network, University of Science and Technology of China, Hefei 230026, China}

\author{Guang-Can Guo}
\affiliation{Laboratory of Quantum Information, University of Science
and Technology of China, Hefei 230026, China}
\affiliation{Anhui Province Key Laboratory of Quantum Network, University of Science and Technology of China, Hefei 230026, China}
\affiliation{CAS Center For Excellence in Quantum Information and Quantum Physics,
University of Science and Technology of China, Hefei, Anhui 230026,
China}
\affiliation{Hefei National Laboratory, University of Science and Technology of China, Hefei 230088, China}

\author{Chang-Ling Zou}
\email{clzou321@ustc.edu.cn}
\affiliation{Laboratory of Quantum Information, University of Science
and Technology of China, Hefei 230026, China}
\affiliation{Anhui Province Key Laboratory of Quantum Network, University of Science and Technology of China, Hefei 230026, China}
\affiliation{CAS Center For Excellence in Quantum Information and Quantum Physics,
University of Science and Technology of China, Hefei, Anhui 230026,
China}
\affiliation{Hefei National Laboratory, University of Science and Technology of China, Hefei 230088, China}

\date{\today}

\begin{abstract}
We propose a Rydberg quantum antenna, consisting of a one-dimensional chain of neutral atoms trapped in a standing-wave optical tweezer, as a chip-interfaced single-photon source. Rydberg blockade induces a single collective excitation in the atomic chain, while its ordered geometry shapes the photon emission into a directional beam, thereby realizing a quantum antenna that emits strictly one photon at a time. Through numerical simulations, we demonstrate single-photon collection efficiencies into a waveguide exceeding 70\% through a high-NA lens with a negligible multiple excitation probability. These performance are robust against probabilistic atom loading in the standing-wave dipole traps, requiring only ten atoms, indicating experimental feasibility. The Rydberg quantum antenna thus offers a unique route toward scalable arrays of high-efficiency, high-purity, and identical single-photon sources interfaced with a photonic chip, addressing key challenges in photonic quantum information technology.
\end{abstract}

\maketitle

\noindent \textit{Introduction.-}
Photonic quantum information processing has emerged as a promising platform for realizing scalable quantum technologies, offering unique advantages such as low decoherence, compatibility with existing optical communication infrastructure, and the potential for chip-scale integration~\cite{OBrien2009,Wang2019,Alexander2025,Wang2025,Zhu2026}. In recent years, significant progress has been made in the development of quantum photonic integrated circuits (QPICs), which enable the integration of essential components, including single-photon sources (SPSs), quantum gates, and detectors, onto a single chip~\cite{Carolan2015,Elshaari2020,Xu2022,Cheng2023,Wang2019}. These advances have opened up new opportunities for applications in quantum computation~\cite{Zhong2020,Madsen2022}, quantum simulation~\cite{Sparrow2018}, and quantum communication~\cite{Yin2017,Stas2022QuantumNetworkNode,Knaut2024TelecomNetwork}.

However, despite the rapid progress in QPIC technology, a major challenge remains in realizing scalable and reliable SPSs that can be seamlessly integrated with photonic chips~\cite{Aharonovich2016}. Existing approaches, such as spontaneous parametric down-conversion (SPDC)~\cite{Kwiat1995}, quantum dots~\cite{Somaschi2016}, and atoms~\cite{Higginbottom2016,Shi2022RydbergSinglePhotonSource}, suffer from limitations in terms of efficiency, indistinguishability, and scalability. SPDC sources rely on probabilistic emission and post-selection, resulting in low efficiency and limited scalability~\cite{Kaneda2019}. Quantum dots offer high single-photon purity but are sensitive to their local environment, leading to spectral inhomogeneity and difficulty in achieving indistinguishable photons from multiple sources~\cite{Senellart2017}. Trapped atoms have demonstrated excellent single-photon properties but require complex laser cooling and trapping setups, making their integration with photonic chips challenging~\cite{Thompson2013,Stephenson2020,Luan2020,Liu2023AtomArrayCavity}. Consequently, there is a strong demand for a scalable and efficient SPS platform that can bridge the gap between atomic systems and photonic quantum technologies.

Here, we propose a novel approach to address the scalability and indistinguishability challenges of SPSs by interfacing Rydberg quantum antenna with photonic chips. Our scheme leverages the collective enhancement of light-matter interactions in one-dimensional atomic lattice confined in a standing-wave optical tweezer~\cite{Chen2024}. Taking advantage of the strong dipole-dipole interactions between Rydberg atoms~\cite{Saffman2010,Dudin2012RydbergExcitations,Browaeys2020}, we demonstrate the possibility of realizing a scalable array of identical SPSs coupled to a single photonic chip, with single photon collection efficiencies exceeding 70\% and a second-order correlation function $g^2$ below 0.01. The proposed Rydberg quantum antenna represents a significant step towards overcoming the long-standing challenges in photonic quantum information processing, offering a promising route for realizing large-scale, high-performance quantum systems with potential applications in quantum computation, metrology, and communication.

\begin{figure*}[t]
\begin{centering}
\includegraphics[width=1.6\columnwidth]{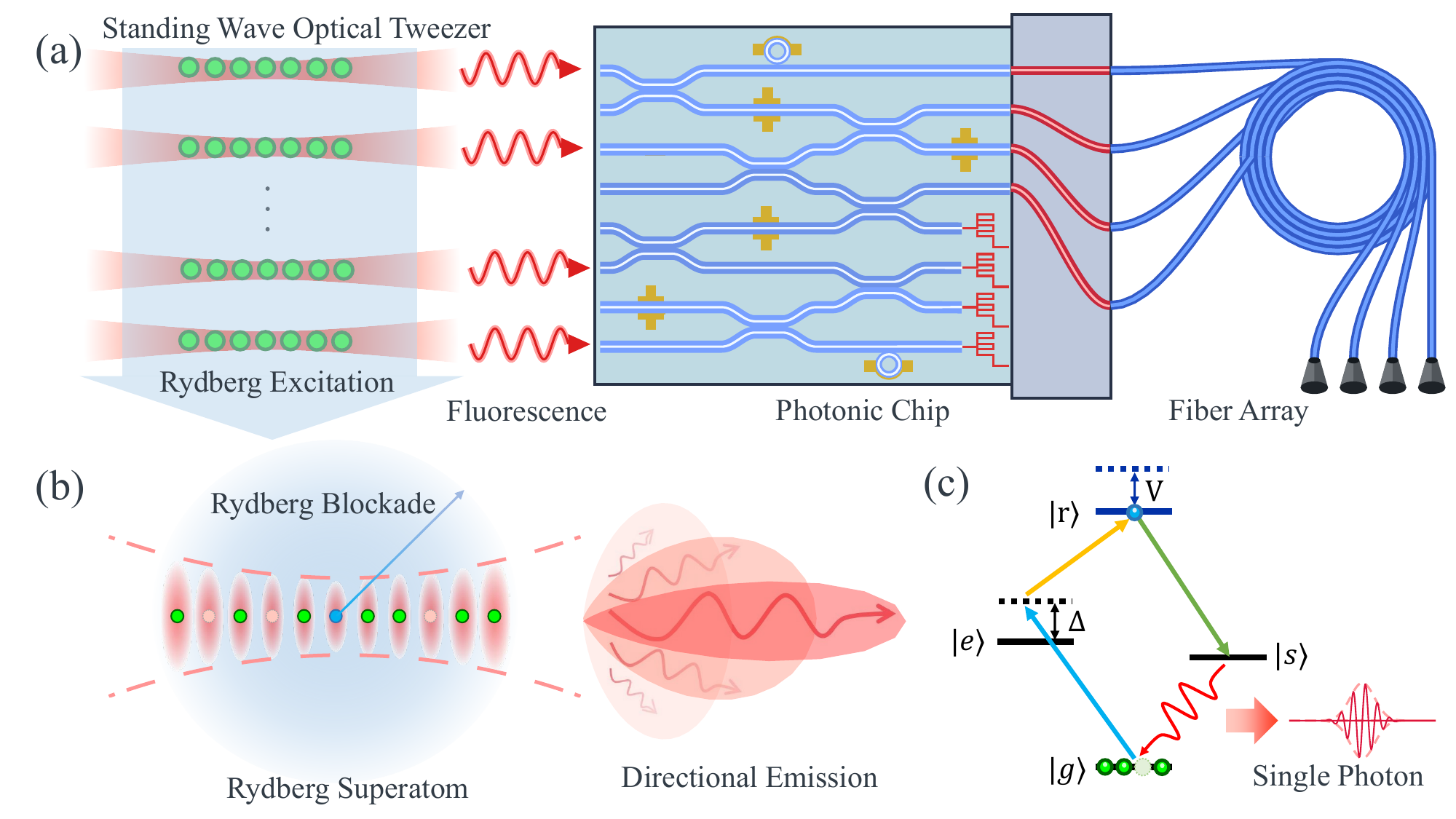}
\par\end{centering}
\caption{\textbf{Rydberg quantum antenna array interfaced with a photonic chip.} (a) Proposed architecture. An array of Rydberg quantum antennas, each a one-dimensional chain of neutral atoms confined in a standing-wave optical tweezer (SWAT), is mapped onto a three-dimensional waveguide array. Single photons are excited and emitted into the waveguide by the antenna under the global excitation lasers. (b) A single Rydberg quantum antenna. The SWAT is formed by retro-reflecting the dipole beam, producing a lattice of traps and atoms are probabilistically loaded into the traps forming a compact atomic chain. Rydberg blockade admits only one collective excitation, and the ordered geometry beams the emitted photon along the chain axis into a single waveguide mode. (c) Atomic level structure. Global drives excite atoms from ground state $\ket{g}$ to $\ket{s}$ through intermediate energy levels $\ket{e}$ and $\ket{r}$, and the directional single photon is through the spontaneous decay of $\ket{s}$.
}
\label{Fig1}
\end{figure*}

\noindent \textit{Principle.-}
Figure~\ref{Fig1}(a) illustrates the proposed architecture, in which an array of Rydberg quantum antennas is directly interfaced with a photonic chip. Each antenna is a one-dimensional chain of neutral atoms confined in a standing-wave optical tweezer (SWAT)~\cite{Wang2023,Chen2024}, and the chains are arranged to match the input facet of a three-dimensional waveguide array on the chip. Following the Volcano architecture for scalable interfacing between single atoms and individual optical channels~\cite{Ma2026}, the waveguide array establishes a one-to-one mapping between each antenna and a dedicated waveguide, so that the single photon emitted by the $j$-th antenna is collected into the $j$-th waveguide and routed, processed, or detected on-chip. Because the emission of each antenna is directional, this mapping turns the free-space collection of atomic fluorescence into an efficient, parallel, and low-crosstalk quantum link, providing a scalable route to large arrays of SPSs.

The core idea of this work is to operate the SWAT-confined atomic chain as a Rydberg quantum antenna, as illustrated in Fig.~\ref{Fig1}(b). Compared with a conventional single-atom emitter or a disordered ensemble, this platform combines four features that are essential for a high-performance, scalable SPS:

(i) \textit{Strong Rydberg interaction.} The sub-wavelength spacing and short Rayleigh length of the traps prepare a compact atomic chain, so that the strong van der Waals interaction admits only a single collective Rydberg excitation, as has been extensively demonstrated with Rydberg superatoms~\cite{Weber2015,Paris-Mandoki2017,Yang2022,Kumlin2023,Shao2024}.

(ii) \textit{Directional emission.} Because the atoms sit at regular, sub-wavelength positions, their emission adds coherently and enables near-unity coupling into a single waveguide mode~\cite{Kim2025MatterWaveRadiance,Rui2020SubradiantMirror,Douglas2026ManyBodyRadiance}.

(iii) \textit{Intrinsically identical photons from global driving.} The emitted photon's frequency is fixed by the driving lasers, so photons produced by different antennas across the array are intrinsically indistinguishable, directly addressing the inhomogeneity bottleneck of solid-state emitters.

(iv) \textit{Reconfigurable and power-efficient trapping.} The SWAT is formed simply by retro-reflecting the dipole beam already present in the optical path, which greatly reduces the required laser power compared with deep single-beam traps that hold many atoms~\cite{Wang2023,Chen2024}. This reflection geometry also makes the tweezer configuration readily reconfigurable, so that the antenna array can be reshaped to match an arbitrary chip layout.

The scheme applies to almost all alkali (e.g., rubidium) and alkaline-earth (e.g., strontium) atoms, with the general atomic level structure shown in Fig.~\ref{Fig1}(c). The atoms are first driven from the ground state $\ket{g}$ to the Rydberg state $\ket{r}$ through either a two-photon or a single-photon excitation process. Owing to the Rydberg interaction, only a single collective excitation is created, then it is coherently transferred to an auxiliary level $\ket{s}$, from which the directional single photon is emitted by relaxation to $\ket{g}$. Recently, the two-dimensional optical tweezer arrays now reach the thousand-site level~\cite{Pause2024,Lin2025,Manetsch2025}, while the three-dimensional waveguide chip of the volcano architecture supports thousand of independent optical channels~\cite{Ma2026,Hu_prep}, indicating that the number of identical SPSs on a single chip can be scalable to $\sim 10^3$ by the Rydberg quantum antenna.

\begin{figure*}[t]
\begin{centering}
\includegraphics[width=1.6\columnwidth]{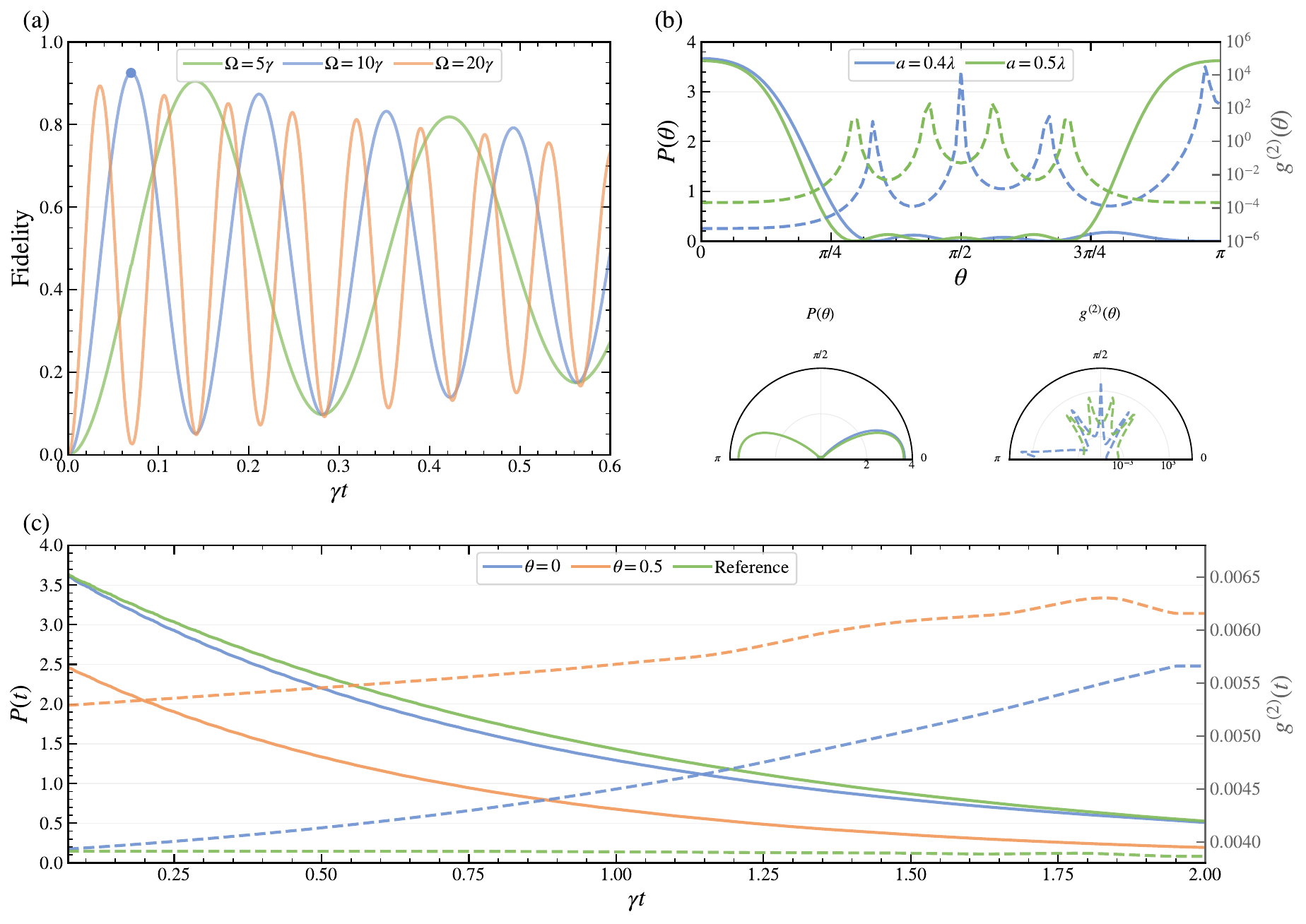}
\par\end{centering}
\caption{\textbf{Directional emission of a single antenna.}  (a) Dynamic evolution of single-excitation fidelity with different driving strengths $\xi_s/\gamma=5$, $10$, and $20$, respectively. The blue dot marks the optimum drive duration $\gamma t=0.07$. (b) Angular distribution $P(\theta)$ (solid lines) and second-order correlation $g^2(\theta)$ (dotted lines) for interatomic spacings $a=0.4\lambda$ and $0.5\lambda$.  (c) Temporal evolution of $P$ and $g^{(2)}$. The green line is the reference line without considering the dipole interaction. The blue and orange lines are the results for different angles $\theta=0$ and $0.5$, respectively, which show that the dipole interaction accelerates the radiation rate. The atom number is $N=5$.
}
\label{Fig2}
\end{figure*}

\smallskip{}
\noindent \textit{Model.-}
At the heart of the Rydberg quantum antenna is the collective, directional emission of a single photon, comprising three stages: preparation of a single collective Rydberg excitation, coherent transfer to the auxiliary state $\ket{s}$, and directional radiation of the $\ket{s}\to\ket{g}$ photon from the SWAT-confined chain. In the interaction picture, the dynamics is governed by the master equation
\begin{eqnarray}
\frac{d\rho}{dt}
&=&
-i\sum_{m=1}^{N}\left[H_{d,m},\rho\right]
-\frac{i}{2}\sum_{m\neq n}^{N}
\left[
V_{mn}\sigma_{rr,m}\sigma_{rr,n},
\rho
\right]
\nonumber\\
&&
+\frac{1}{2}\sum_{m=1}^{N}
\left[
\gamma_{gr}\mathcal{L}\left(\sigma_{gr,m}\right)
+\gamma_{sr}\mathcal{L}\left(\sigma_{sr,m}\right)
\right]
\nonumber\\
&&
-i\sum_{m\neq n}^{N}
\left[
\Delta_{mn}
\sigma_{gs,m}^{\dagger}
\sigma_{gs,n},
\rho
\right]
+\frac{1}{2}
\int
\mathcal{L}
\left(
{S}(\theta,\phi)
\right),
\label{Eq1}
\end{eqnarray}
where the driving field is
$
H_{d,m}
=
\xi_e
\sigma_{gr,m}^{\dagger}
e^{i{\textbf{k}_e}\cdot{\textbf{r}_m}}
+
\xi_s
\sigma_{sr,m}^{\dagger}
e^{i{\textbf{k}_s}\cdot{\textbf{r}_m}}
+\mathrm{h.c.}
$
with the wave vector ${\textbf{k}_e}\left({\textbf{k}_s}\right)$ and the amplitude $\xi_e\left(\xi_s\right)$, ${\textbf{r}_{m}}$ represents the position of the $m$-th atom, and $\sigma_{e_{1}e_{2},m}=\left|e_{1}\right\rangle _{m}\left\langle e_{2}\right|$ is the Pauli operator for the atom with $e_{1\left(2\right)}=g,~e,~r,~s$. $V_{mn}=C_{6}/\left|{\textbf{r}_{m}}-{\textbf{r}_{n}}\right|^{6}$ is the Rydberg interaction, which causes the atomic chain to form Rydberg superatoms. $\mathcal{L}\left(o\right)=2o\rho o^{\dagger}-o^{\dagger}o\rho-\rho o^{\dagger}o$ is the Lindblad superoperator, and the middle two terms refer to the dissipation of atoms, with $\gamma_{gr(sr)}$ denotes the decay from Rydberg states to $\ket{g(s)}$. Throughout this work, we set $\gamma_{gr}=\gamma_{sr}=\gamma$ as the common single-atom spontaneous decay rate. The terms proportional to $\Delta_{mn}$ in the master equation account for the dipole-dipole interactions and the collective spontaneous emission can be expressed as the directed-detection
jump operators
\begin{equation}
{S\left(\theta,\phi\right)}=\sqrt{\gamma D\left(\theta,\phi\right)d\Omega}\sum_{m=1}^{N}e^{-ik_{0}{\textbf{R}\left(\theta,\phi\right)}\cdot{\textbf{r}_{m}}}\sigma_{gs,m},
\end{equation}
which indicates a photon emitted into the solid angle $d\Omega$ in the direction ${\textbf{R}\left(\theta,\phi\right)}$. Here, $D\left(\theta,\phi\right)=3/\left(8\pi\right)\left\{ 1-\left[{\textbf{d}}\cdot{\textbf{R}\left(\theta,\phi\right)}\right]^{2}\right\}$ is
the single-atom dipole radiation pattern aligned along ${\textbf{d}}$. The set ${\hat S(\theta,\phi)}$ provides an equivalent unraveling of the collective dissipator, since $\int d\Omega S^\dagger\hat S=\sum_{m,n}\Gamma_{mn}\sigma_{sg,m}\sigma_{gs,n}$.

In an ideal experiment, the excitation and transfer prepare the chain in the single-excitation spin-wave state $\left|\psi\right\rangle =\frac{1}{\sqrt{N}}\sum_{m=1}^{N}e^{i\left(m-1\right)\delta\theta}\left|g_1\cdots s_m\cdots g_N\right\rangle$, with atoms located at ${\textbf{r}_{m}}=\left(m-1\right)\left(a\lambda_{0}\right){\textbf{e}_{z}}$, where $a$ is the interatomic separation in units of the emitted photon wavelength $\lambda_{0}$ and $\delta\theta=\left(a\lambda_{0}\right)\left({\textbf{k}_e}-{\textbf{k}_s}\right)\cdot{\textbf{z}}$ is the imprinted phase gradient. The far-field intensity $P\left(\theta,\phi\right)d\Omega=\left\langle \psi\right|{S\left(\theta,\phi\right)}^{\dagger}{S\left(\theta,\phi\right)}\left|\psi\right\rangle$ then evaluates as
\begin{equation}
P\left(\theta,\phi\right)d\Omega =\left(\frac{\sin N\phi_s}{\sin \phi_s}\right)^{2}\frac{\gamma D\left(\theta,\phi\right)d\Omega}{N}
\label{IdealEmission}
\end{equation}
with $\phi_s=\pi a\cos\left(\theta\right)-\delta\theta/2$. The emission is thus governed by the array factor $|\sin N\phi_s/\sin\phi_s|^2$, in direct analogy to the radiation pattern of a phased-array antenna.

To beam the photon toward one end of the chain ($\theta=0$), we choose circularly polarized dipoles $\textbf{d}=( \textbf{e}_x+i \textbf{e}_y)/\sqrt2$, giving $D(\theta,\phi)=\tfrac{3}{8\pi}[1+\cos^2\theta]$, and match the drive phase to $\delta\theta=2\pi a$. A straightforward analysis shows that unidirectional emission requires a sub-wavelength spacing $a<1/2$: for $a>1/2$ a backward diffraction order appears and the forward main lobe carries less than $50\%$ of the emission, whereas for $a<1/2$ over $90\%$ is beamed forward, with the main-lobe width narrowing as $1/N$.

\begin{figure*}[t]
\begin{centering}
\includegraphics[width=1.6\columnwidth]{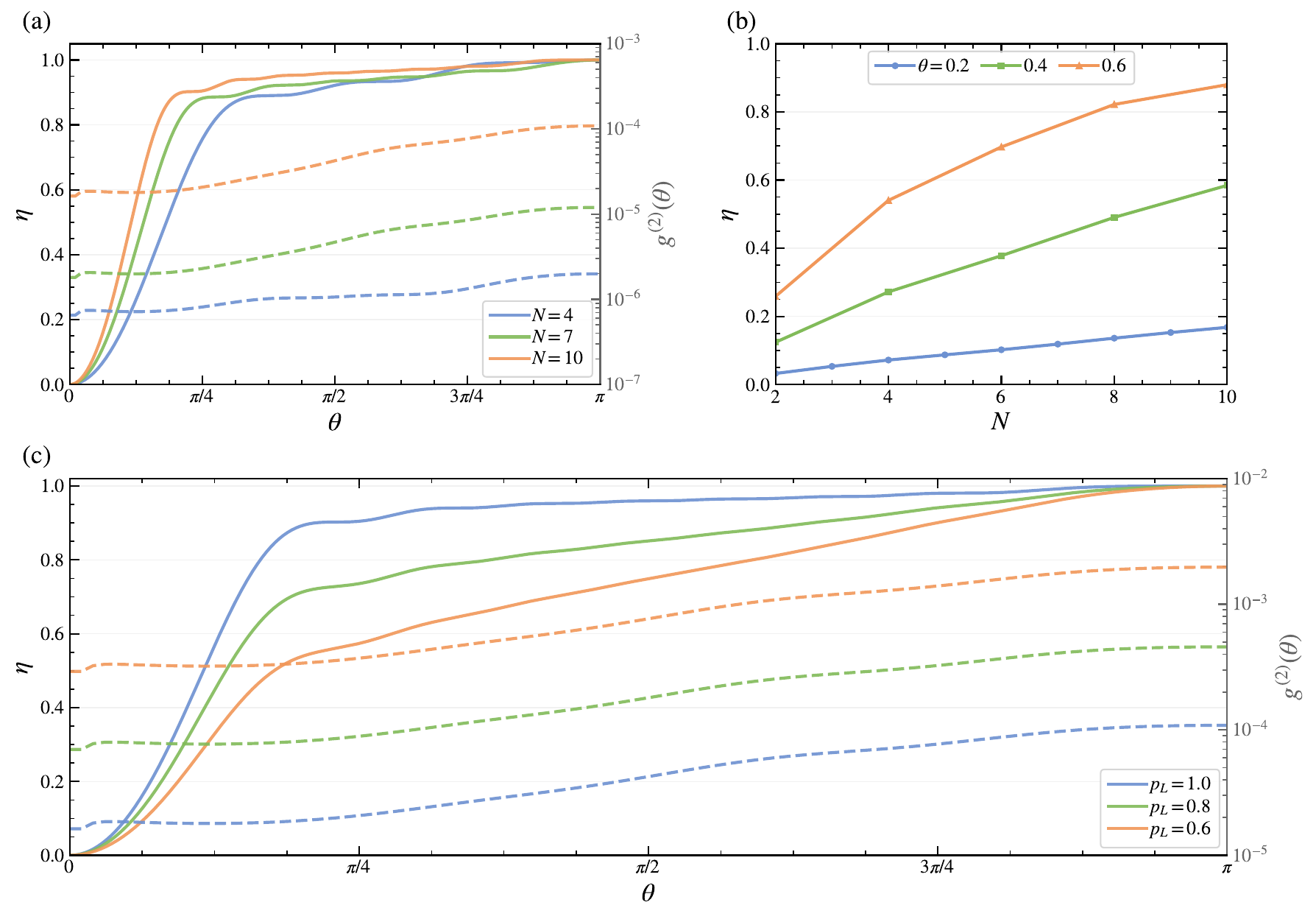}
\par\end{centering}
\caption{\textbf{Collection efficiency.}  (a) The collection efficiency $\eta$ (solid) and second-order correlation function $g^{(2)}$ (dotted) versus the collection half-angle $\theta$ for atom numbers $N=4,7$, and $10$, respectively. (b) $\eta$ as functions with the atom number $N$ at fixed collection half-angles. (c) The mean performances of the antenna for atom loading probabilities $p_L=1,\,0.8,\,0.6$ at  $N=10$. Each data point for $p_L<1$ are averaged over 500 random loading configurations.
}
\label{Fig3}
\end{figure*}

\smallskip{}
\noindent \textit{Directional single-photon emission.-}
The Eq.~\ref{IdealEmission} describes an ideal antenna. To assess its performance as a practical SPS, we now evaluate the emission under three dominant imperfections: First, the excitation of multiple Rydberg atoms due to imperfect Rydberg blockade, and this multi-photon contamination is quantified by the second-order correlation $g^{(2)}$, which vanishes for an ideal single excitation and grows with the dominant residual double excitation. Second, the directional emission is not perfectly collimated, so the fraction of photons collected within a finite aperture is quantified by the collection efficiency $\eta(\theta)$. Third, the SWAT loads atoms probabilistically, so a valid antenna must tolerate randomly missing sites, and we therefore assess the robustness of both $\eta$ and $g^{(2)}$ against a reduced loading probability. Therefore, we characterize these effects by numerically solving the master equation [Eq.~(\ref{Eq1})], with the parameter $C_6=8\,\mathrm{GHz}/\mathrm{\mu m}^6$ and atomic spacing $0.5\lambda$. In Fig.~\ref{Fig2}, we first evaluate the antenna performance with a fully filled SWAT, with the number of filled site $N=5$.

We begin with the preparation of the collective excitation, which sets the ultimate single-photon purity of the antenna. Figure~\ref{Fig2}(a) shows the single-excitation fidelity as a function of the pulse duration $\gamma t$ for several drive strengths $\xi_s$. The dynamics reflect a competition between two processes: a stronger drive speeds up the excitation but enhances the residual double excitation allowed by the imperfect blockade, whereas a longer pulse increases the photon loss through dissipation during the evolution. As a result, an optimal drive $\xi_s/\gamma=10$ achieves the highest fidelity at $\gamma t\approx0.07$ by balancing multiple excitation and photon loss. The subsequent transfer from $\ket{r}$ to the auxiliary state $\ket{s}$ involves no multiple excitations and can be driven strongly, we treat it as an instantaneous, lossless mapping and take this optimal drive for the following analysis in Figs.~\ref{Fig2}(b) and (c).

Figure~\ref{Fig2}(b) shows both the radiation intensity $P(\theta)$ and the angle-resolved $g^{(2)}(\theta)$. As predicted in the ideal case [Eq.~(\ref{IdealEmission})], for $a=0.5$, the emission is symmetric with respect to the chain. Reducing the spacing to $a=0.4$ breaks this symmetry: most of the emission is now beamed toward one side, in agreement with the unidirectionality condition $a<1/2$ derived above. The corresponding correlation $g^{(2)}(\theta)$ oscillates with angle but remains far below unity at major emission directions for both $a=0.4$ and $0.5$, confirming the antibunched, single-photon character of the emission.

Figure~\ref{Fig2}(c) shows the temporal evolution of the emitted photon profile and of $g^{(2)}$. To isolate the role of the collective coupling, we compare the full dynamics (blue line) with a reference case in which the resonant dipole-dipole interaction is switched off (green line). Including the interaction accelerates the decay of the excitation, reflecting the collectively enhanced emission of the ordered chain. The same speedup persists when the emission is resolved at other angles (e.g., $\theta=0.5$). Throughout the entire radiative process, $g^{(2)}$ stays below $10^{-3}$, so the chain preserves both its unidirectionality and its single-photon statistics as it radiates.

Having established the angular and temporal behavior of a single antenna, we now study its performance in terms of the collection efficiency $\eta$. Since three-dimensional waveguides support near-Gaussian fundamental modes~\cite{Ma2026}, $\eta$ is defined as the fraction of emission funneled into a forward cone of half-angle $\theta$ for simplicity, with $\eta=\int_{0}^{\theta}P\left(\theta^{'}\right)\sin\left(\theta^{'}\right)d\theta^{'}/\int_{0}^{\pi}P\left(\theta^{'}\right)\sin\left(\theta^{'}\right)d\theta^{'}$. Here, $\theta$ maps directly onto the numerical aperture of the collection optics through $\mathrm{NA}=\sin\theta$. A more rigorous treatment can be obtained by free-space-beam quantum electrodynamics, in which atomic thermal motion effects can be included~\cite{Chen2024OL}.

Figure~\ref{Fig3}(a) shows $\eta(\theta)$ and the averaged $g^{(2)}$ as functions of the collection angle for different atom numbers $N$. The directionality is excellent: even with only four atoms, $\eta$ saturates toward unity once the cone exceeds $\theta\approx1$, while the averaged $g^{(2)}$ remains far below $10^{-4}$ over the entire angular range, consistent with the angle-resolved result of Fig.~\ref{Fig2}(b). The directionality further improves with atom number, as confirmed in Fig.~\ref{Fig3}(b) at fixed collection angle. For $N=10$, we achieve $\eta=90\%$ within a cone of $\theta=0.6$ ($\mathrm{NA}\approx0.56$), and still $\eta=60\%$ within a tighter cone of $\theta=0.4$ ($\mathrm{NA}\approx0.39$), while the averaged $g^{(2)}$ stays below $10^{-3}$. This trend reflects the narrowing of the main lobe as $1/N$ set by the array factor, so that larger antennas beam their photons into progressively smaller apertures.

Finally, because the SWAT loads atoms probabilistically, we assess the robustness against randomly missing sites in Fig.~\ref{Fig3}(c). For $p_L<1$, each of the ten sites is independently occupied with probability $p_L$, and the reported $\eta$ and $g^{(2)}$ are averages over 500 random loading configurations. For a chain with 10 sites and loading probabilities of $p_L=0.8$ and $0.6$, both $\eta$ and $g^{(2)}$ remain essentially unchanged, demonstrating that the antenna performance is resilient to atom loss. This robustness is a direct consequence of the collective nature of the emission, which depends on the ordered ensemble rather than on any individual site being occupied.

\smallskip{}
\noindent\textit{Conclusion}
We have proposed and analyzed a Rydberg quantum antenna for realizing high-performance arrays of directional SPSs interfaced with photonic chips. Each antenna is built by extending a single-atom tweezer with a retro-reflected dipole laser. The atoms confined in the SWAT are prepared and excited through global driving lasers, so that the antenna array can be scaled up to thousands whose emitted photons are intrinsically identical~\cite{Pause2024,Lin2025,Manetsch2025}, directly addressing the indistinguishability and scalability bottlenecks of SPSs. Moreover, the Rydberg quantum antenna may be generalized into a matter qubit that stores and processes quantum states, and hybrid entangling gates between single atoms and antennas are worth further exploration. Moreover, as the time reversal of emission, the antenna can act as a quantum memory~\cite{Manzoni2018}, providing a key ingredient for entanglement distribution and positioning the antenna array as a resource for photonic quantum computation, metrology, and networking. Therefore, our work appeals for dedicated experimental efforts on the practical limitations, such as the impact of trap-depth inhomogeneity and dipole laser field distortion and the decoherence channels of the quantum information stored in the SWAT.

\smallskip{}
\begin{acknowledgments}
This work was funded by the National Key R\&D Program (Grant No.~2021YFA1402004), and the National Natural Science Foundation of China (Grants No.~92265210 and 92465201), and the Shenzhen International Quantum Academy (Grant No.SIQA2025KFKT02). This work was also supported by the Fundamental Research Funds for the Central Universities and USTC Research Funds of the Double First-Class Initiative. The numerical calculations in this paper have been done on the supercomputing system in the Supercomputing Center of USTC. This work was partially carried out at the USTC Center for Micro and Nanoscale Research and Fabrication.
\end{acknowledgments}

\end{document}